\documentclass[sigconf,screen,authorversion]{acmart}

\usepackage[utf8]{inputenc}        
\usepackage[nameinlink]{cleveref}  
\usepackage{enumitem}              

\newcommand{\enquote}[1]{``#1''}  
\newcommand{\checklistitem}[1]{\\[\smallskipamount] 
#1}

\AtBeginDocument{%
  }

\copyrightyear{2025} 
\acmYear{2025} 
\setcopyright{rightsretained} 
\acmConference[SIGCSE TS 2025]{Proceedings of the 56th ACM Technical Symposium on Computer Science Education V. 1}{February 26--March 1, 2025}{Pittsburgh, PA, USA}
\acmBooktitle{Proceedings of the 56th ACM Technical Symposium on Computer Science Education V. 1 (SIGCSE TS 2025), February 26--March 1, 2025, Pittsburgh, PA, USA}
\acmDOI{10.1145/3641554.3701976}
\acmISBN{979-8-4007-0531-1/25/02}

\newcommand{\thewu}{\textcolor[HTML]{e41a1c}{THE-WU}}
\newcommand{\qswu}{\textcolor[HTML]{377eb8}{QS-WU}}
\newcommand{\edu}{\textcolor[HTML]{059d00}{EDU}}
\newcommand{\gci}{\textcolor[HTML]{984ea3}{GCI}}
\newcommand{\theee}{\textcolor[HTML]{e17000}{THE-EE}}

\begin{document}

\title{Cybersecurity Study Programs: What's in a Name?}

\author{Jan Vykopal}
\orcid{0000-0002-3425-0951}
\affiliation{
    \institution{Masaryk University}
    \department{Faculty of Informatics}
    \city{Brno}
    \country{Czech Republic}
}
\email{vykopal@fi.muni.cz}

\author{Valdemar Švábenský}
\orcid{0000-0001-8546-280X}
\affiliation{
    \institution{Kyushu University}
    \department{Faculty of Information Science and Electrical Engineering}
    \city{Fukuoka}
    \country{Japan}
}
\email{valdemar.research@gmail.com}

\author{Michael Tuscano Lopez II}
\orcid{0009-0001-8092-2587}
\affiliation{
    \institution{Ateneo de Manila University}
    \department{Department of Information Systems and Computer Science}
    \city{Quezon City}
    \country{Philippines}
}
\email{michael.lopez@student.ateneo.edu}

\author{Pavel Čeleda}
\orcid{0000-0002-3338-2856}
\affiliation{
    \institution{Masaryk University}
    \department{Faculty of Informatics}
    \city{Brno}
    \country{Czech Republic}
}
\email{celeda@fi.muni.cz}

\begin{abstract}
Improving cybersecurity education has become a priority for many countries and organizations worldwide. Computing societies and professional associations have recognized cybersecurity as a distinctive computing discipline and created specialized cybersecurity curricular guidelines. Higher education institutions are introducing new cybersecurity programs, attracting students to this expanding field. In this paper, we examined 101 study programs across 24 countries. Based on their analysis, we argue that top-ranked universities have not yet fully implemented the guidelines and offer programs that have \enquote{cyber} in their name but lack some essential elements of a cybersecurity program. In particular, most programs do not sufficiently cover non-technical components, such as law, policies, or risk management. Also, most programs teach knowledge and skills but do not expose students to experiential learning outside the traditional classroom (such as internships) to develop their competencies. As a result, graduates of these programs may not meet employer expectations and may require additional training. To help program directors and educators improve their programs and courses, this paper offers examples of effective practices from cybersecurity programs around the world and our teaching practice.
\end{abstract}


\keywords{cyber security, curricular guidelines, CSEC2017, CC2020, higher education, university programs, global perspective}

\maketitle

\section{Introduction} \label{sec:introduction}

Due to the global prevalence of cyber crime and its impacts~\cite{Chen2023}, cybersecurity is of utmost importance in a wide variety of sectors, from business through healthcare to national security. As a result, enhancing cybersecurity education is a strategic goal defined in the policies of numerous countries across the world~\cite{AlDaajeh2022}, aiming to address the growing demand for cybersecurity professionals~\cite{Sui2024}. 

Higher education institutions have been responding to this demand by establishing cyber study programs to attract students to this field, which is expected to continue growing. Subsequently, educators have been sharing their insights and perspectives to encourage dialogue around cybersecurity curricula and facilitate their improvements. Particularly at computing education venues like SIGCSE, this topic has been of interest in recent years~\cite{Crabb2024, Blaine2021, Asghar2020, Weiss2021}. 

\subsection{Situating Cybersecurity Within Computing}
\label{subsec:intro-cybersecurity}

\textit{Computing} includes \textit{\enquote{computer science and related disciplines, such as computer engineering, information systems, information technology, software engineering, cybersecurity, and data science}}~\cite{Guzdial2018}. Curricular efforts to define the content of computing degree programs date from at least the 1960s, and the evolution of cybersecurity coverage within these curricula is notable, as described by~\cite{Svabensky2022thesis}. The influential 1968 ACM recommendations~\cite{Atchison1968} do not mention the term \enquote{security} at all. Another seminal text from 1989~\cite{Denning1989} makes only a brief reference to \enquote{secure computing}. Today, ACM/IEEE Computing Curricula 2020 (CC2020)~\cite{cc2020} recognize cybersecurity as an essential component of educating computing professionals. Moreover, specialized cybersecurity curricular guidelines have emerged in recent years~\cite{Parrish2018, Mouheb2019}, most notably:
\begin{itemize}[leftmargin=22pt]
    \item ACM/IEEE/AIS SIGSEC/IFIP JTF Cybersecurity Curricula -- \textbf{CSEC2017}~\cite{csec2017} and its adaptation for community colleges, Cybersecurity Curricular Guidance -- \textbf{Cyber2yr2020}~\cite{cyber2yr2020},
    \item NIST National Initiative for Cybersecurity Education -- \textbf{NICE} (USA)~\cite{nice2020},
    \item NCSE Cyber Security Body of Knowledge -- \textbf{CyBOK} (UK)~\cite{cybok2021},
    \item ENISA European Cybersecurity Skills Framework -- \textbf{ECSF} (EU)~\cite{ecsf2022}.
\end{itemize}

Cybersecurity is a multidisciplinary field encompassing various computing domains, including secure programming, network defense, and operating system administration, along with non-technical aspects~\cite{Svabensky2022thesis}. To become proficient in these subjects, students require theoretical foundations and practical experience with relevant tools and processes~\cite{csec2017}. Delivering this experience necessitates realistic learning environments and current, relevant tasks. However, since cybersecurity is an ever-evolving field, tools and procedures that worked yesterday might be obsolete today. Fortunately, there are modern opportunities for students to enhance their cybersecurity knowledge and skills~\cite{weitl2023systematic, Crabb2024, Svabensky2020what}.

\subsection{Current Motivating Context}

While understanding cybersecurity concepts is an essential starting point for prospective cybersecurity experts, it alone is not fully sufficient~\cite{Mouheb2019}. Cybersecurity professionals also require hands-on experience with applying cybersecurity skills in practical contexts. As early as 2014, it was observed that many university graduates were unprepared for cybersecurity roles due to a lack of industry-required practical skills~\cite{Conklin2014}. This gap between academic preparation and workplace requirements has been pointed out again in the CSEC2017 guidelines~\cite{csec2017}. Yet, these concerns still remain relevant in 2024: Crabb et al.~\cite{Crabb2024} observed that curricular recommendations for universities emphasize knowledge acquisition and understanding of concepts, whereas the industry expects job candidates with higher-level application skills. Lastly, \citet{Sui2024} raises similar concerns about \enquote{noncompliance with international standards}.

\subsection{Goals and Scope of This Paper}

\begin{figure}[!h]
\centering
\includegraphics[width=0.8\linewidth]{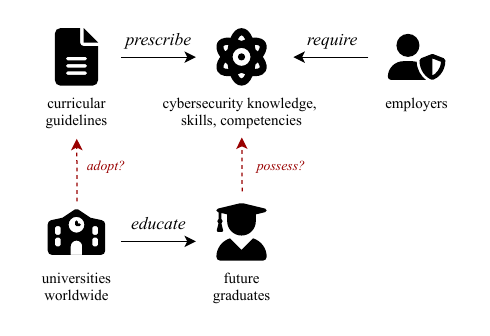}
\caption{Cybersecurity curricular guidelines define knowledge, skills, and competencies required by employers. But do universities adopt these guidelines in their degree programs?}
\Description{This diagram illustrates the relationship between cybersecurity knowledge, skills, and competencies, which are required by employers and prescribed by curricular guidelines, to their adoption by universities and subsequently the mastery of these topics by graduates.}
\label{fig:intro}
\end{figure}

This paper examines \emph{whether cybersecurity study programs offered by leading universities incorporate the elements recommended in curricular guidelines and valued by employers} (see \Cref{fig:intro}). We analyzed cybersecurity study programs offered by 101 universities and educational institutions globally, focusing on descriptions of these programs and offered courses. Therefore, we aim to address overarching issues beyond individual course details. Specifically, we closely inspected \emph{i)} the inclusion of non-technical aspects, such as law, policy, human factors, ethics, and risk management, and \emph{ii)} the integration of experiential learning within these study programs.

Based on our findings and teaching experience, we discuss the current state of cybersecurity education and summarize recommendations and best practices.
We believe our work is beneficial for study program directors, coordinators, and educators who design and deliver the programs and courses.

\subsection{Paper Outline}
\Cref{sec:related} summarizes previous related academic research.
\Cref{sec:survey} details the methods used to select and analyze the offered study programs.
\Cref{sec:findings} presents the findings and highlights examples of best practices.
\Cref{sec:checklist} offers a checklist for directors of current and prospective cybersecurity programs.
\Cref{sec:conclusion} concludes the paper and provides recommendations for future improvements.

\section{Prior and Related Work} \label{sec:related}

Since our paper examines entire study programs, we review other publications of a similar type. This means we focus on papers that also deal with university programs or broader curricular issues (in \Cref{subsec:related-work-programs}), not papers about a single course or an isolated cybersecurity exercise. \Cref{subsec:related-work-job} follows up with a discussion of skills that cybersecurity graduates need for the job market.

\subsection{Cybersecurity Study Programs}
\label{subsec:related-work-programs}

The most recent and related paper to ours is a review of cybersecurity education in the USA~\cite{Crabb2024}. The authors analyzed descriptions of cybersecurity programs at 100 institutions -- Centers of Academic Excellence (CAE) in Cybersecurity. They concluded that these institutions should ensure their curricula match the needs of industry, so that their graduates are work-ready specialists. 

While \citet{Crabb2024} focused on the USA, our analysis is global, including 68 institutions outside the USA. 
Finally, \citet{Crabb2024} did not publish a dataset or supplementary materials providing details about the analyzed programs. As a result, we are unable to report the overlap in terms of which programs they analyzed are also included in our study. We can only report that our dataset contains 13 programs of the CAE-designated institutions.

Next, \citet{Blaine2021} described the evolution of a cyber science study program at the US Military Academy. The study program aims at undergraduates and is consistent with various curricular guidelines, including CSEC2017~\cite{csec2017}. It consists of core foundational courses, such as programming and networking, and elective specializations, such as cybersecurity, cyber operations, or cyber-physical systems. Based on their experience, the authors provide recommendations to faculty members at other institutions that consider adding a new cyber major. For example, they suggest \enquote{an incremental approach [\dots] by progressively introducing electives to existing programs, such as [computer science]}.

\citet{Asghar2020} reflected on a cybersecurity master degree program in New Zealand. They mapped the content of the courses to a competency-based framework~\cite{Parrish2018}, demonstrating the coverage of adversarial aspects as well as interdisciplinary elements. Apart from describing the courses and graduate profiles, they also shared the experience with institutional requirements, such as quality assurance and program evaluation. Similarly to \citet{Blaine2021}, their experience may be valuable for educators who intend to establish a new degree program at their university.

\citet{daVeiga2021} sought to create a cybersecurity curriculum for universities by examining the current best practices. They focused on their national context of South Africa, concluding that \enquote{South African universities do not have a formal undergraduate course in cybersecurity} (as of 2021). Therefore, they selected five high-ranked universities (three in Australia and two in the UK) with cybersecurity degree programs, and complemented their review with seven academic survey papers, resulting in 49 cybersecurity modules. The most commonly occurring ones featured skills regarding networks, cryptography, and forensics.

Overall, while the prior publications reviewed above examine relevant curricular initiatives, to the best of our knowledge, no paper provides a global perspective as broad as our analysis.

\subsection{Cybersecurity Job Market}
\label{subsec:related-work-job}

\citet{Graham2023} analyzed 17,929 online job advertisements for cybersecurity positions. Apart from the requirements on technical skills, which were often focused on data (data protection, loss prevention, and privacy), the job postings strongly emphasized non-technical skills, including communication, teamwork, and problem-solving. This is consistent with earlier work by \citet{Jones2018} who interviewed 44 cyber professionals on the most important skills that cybersecurity students should learn in school. The technical skills included, for example, understanding how network traffic flows and how to harden operating systems. The non-technical skills included mainly communication, presentation skills, and collaboration.

In a related work, \citet{Ozyurt2024} analyzed 9,407 cybersecurity job postings. The advertised job roles were divided into ten categories (e.g., engineer, analyst, or manager), with each category further subdivided into specific job titles. In addition, automated topic modeling identified 23 topic categories (e.g., security operations, risk management, and business/customer services) within the job postings. The mapping between the job roles and the topics revealed which types of skills are more relevant for which types of positions, highlighting the spectrum between more technically-oriented and more human/organization-oriented roles.

Generally, the analysis of the current job market proves the importance of both technical and non-technical competencies, as well as the value of hands-on skills. This is strongly aligned with the cybersecurity curricular guidelines that we use as a baseline. 

\section{Global Survey of Cyber Programs} \label{sec:survey}

\begin{figure}[t]
\centering
\includegraphics[width=1.00\linewidth]{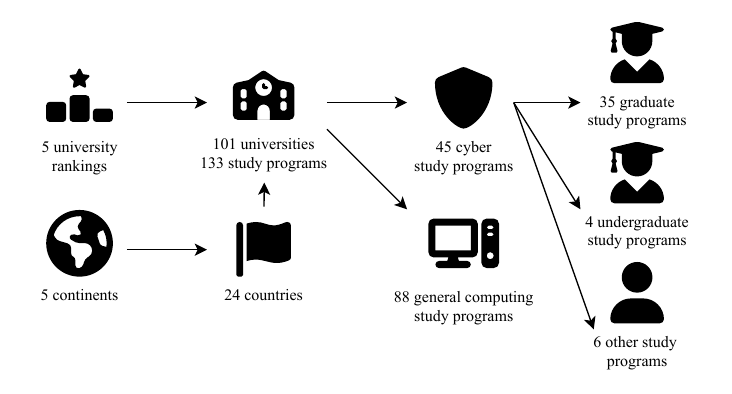}
\caption{Overview of the explored study programs.}
\Description{This diagram illustrates how we used 5 different university rankings to identify 133 study programs at 101 universities from 24 countries on 5 continents. Among the 133 study programs, 45 were cyber programs and 88 were general computing programs. Among the 45 cyber programs, 35 were graduate, 4 undergraduate, and 6 other.}
\label{fig:study-diagram}
\end{figure}

In this paper, we seek to understand how the CSEC curricular guidelines~\cite{csec2017}, which were published in 2017, have been adopted in higher education. To achieve this goal, we review study programs from 101 universities and present our position through the subsequent data analysis. \Cref{fig:study-diagram} shows the overview of our approach. 

The authors of this paper have extensive experience with cybersecurity education -- both with teaching and publishing research in this field. Two co-authors are associate professors at a computer science department of a large university, which has a dedicated 3-year bachelor's program in cybersecurity and 2-year master's programs in information security and cybersecurity management. They have been teaching cybersecurity at a university level since 2015 and are interested in how it is taught worldwide. \emph{The authors do not come from universities whose programs were examined.}

\subsection{Selection of Universities}

We surveyed top universities according to five different internationally recognized rankings (color-coded for readability in \Cref{tab:countries}):

\begin{enumerate}[leftmargin=22pt]
    \item top 50 from Times Higher Education World University Rankings 2024 by the subject of computer science~\cite{the-world} (\thewu),
    \item top 50 from QS World University Rankings 2024 by the subject of computer science and information systems~\cite{qs-world} (\qswu), 
    \item top 50 from EduRank Best Universities for Cybersecurity 2024 in the World~\cite{edurank} (\edu),
    \item top 31 universities from top 10 countries in Global Cybersecurity Index 2020 (\gci) capturing \enquote{the commitment of countries to cybersecurity at a global level}~\cite{gci},
    \item top 15 from Times Higher Education Emerging Economies University Rankings 2022~\cite{the-emerging} (\theee).
\end{enumerate}
 
The rationale behind this selection is an attempt to include not only the top universities in computer science, which are predominantly located in the USA, but also those focusing on cybersecurity and/or outside the USA. 
To complement the two traditional rankings (\thewu\ and \qswu), we use \edu, which relies only on metrics of research performance (OpenAlex citation database), non-academic prominence (backlinks on the Internet), and alumni score (page views at Wikipedia)~\cite{edurank}. We added the top 50 universities in the world and  
at least two highest-ranked universities in \edu\ from each country in the top 10 countries in the \gci\ except the USA (\#1 in \gci)\footnote{If we had already added two universities from a country from \gci\ based on \thewu\ or \qswu, we added a third based on \edu. There were 14 countries ranked \#2 to \#10 in \gci; for three of them we added the third university, which resulted in 31 in total.}.
Finally, to ensure truly global coverage, we searched for cybersecurity programs also at universities from emerging economies (\theee) that may be motivated to keep up with the developed countries and established universities.

\subsection{Extraction of Study Programs' Information}

Two authors conducted the survey of study programs, with the first author double-checking and confirming the collected information to ensure its validity. 

For each university, we searched their respective websites for any offered cybersecurity study programs, except Ph.D. (doctoral). If the said institution did not offer a program with the word \enquote{cyber} in its name, then we resorted to looking for their post-graduate degrees that are related to computer security/science. Examples of these include \textit{Master in Information Security} or \textit{Master in Informatics}. Undergraduate and associate degrees in computer science were not considered in the analysis. In sum, we tried to exhaust all computing-related post-graduate degrees to be included in our dataset as proof we did not omit to examine the offerings of a given institution. 

For all programs having \enquote{cyber} in their name, we manually collected their type, duration, links to their curricula, and study or course catalogs (if available). Then, we inspected the catalogs and course descriptions to find out \emph{i)} whether the core courses or core electives cover the CSEC2017 areas of human, organizational, and societal security (see \Cref{subsec:non-technical}); and \emph{ii)} hands-on learning components (see \Cref{sec:hands-on}), namely: requirements for completing internships (work placements); a semester/pre-capstone project, the final capstone project, or final thesis.

\subsection{Limitations of the Survey Approach}

There are two main limitations that could influence the findings. First, if a university or a country is not present in any of the rankings (\thewu, \qswu, \edu, \gci, or \theee), its cybersecurity program will not appear in our analysis, even though it may be of high quality. 
Second, if the website of the study program has not been updated recently by the university, the information may not fully reflect the latest status of the program.

\section{Findings and Examples of Good Practice} \label{sec:findings}

This section summarizes the survey results and highlights good practices. The dataset with all collected information about the study programs and a Python notebook for its processing is available~\cite{dataset}.

\subsection{Composition of Discovered Programs}

A total of 133 study programs at 101 universities from 24 countries (see~\Cref{tab:countries}) were enumerated.  
The Spearman correlation among \thewu\ and \edu\ rankings of all universities is $0.68, p<0.001$. The correlation among \qswu\ and \edu\ is $0.65, p<0.001$. This means that universities generally higher in \thewu\ or \qswu\ do not necessarily merit a similar ranking in \edu, supporting our decision to include multiple rankings in the analysis.
 
\begin{table}[h!]
\caption{Breakdown of 101 examined institutions per country.}
\centering
\footnotesize
\begin{tabular}{l|c|l}
\textbf{Country} & \textbf{Institutions} & \textbf{Source ranking(s)} \\ \hline 
USA & 33 & \thewu, \qswu, \edu \\ 
China (Mainland) & 12 & \thewu, \qswu, \edu, \theee \\
United Kingdom & 6 & \thewu, \qswu, \edu, \gci \\
Canada & 5 & \thewu, \qswu, \gci \\
France & 4 & \thewu, \qswu, \gci \\
Russian Federation & 4 & \gci, \theee \\
Australia & 3 & \qswu, \edu \\ 
China (Hong Kong) & 3 & \thewu, \qswu, \edu \\
Korea (Rep. of) & 3 & \thewu, \qswu, \gci\\ 
Estonia & 2 & \gci\\
Germany & 2 & \thewu, \qswu \\
India & 2 & \gci \\ 
Japan & 2 & \thewu, \qswu, \gci \\
Lithuania & 2 & \gci \\ 
Malaysia & 2 & \gci \\
Netherlands & 2 & \thewu, \qswu \\
Saudi Arabia & 2 & \gci, \theee \\
Singapore & 2 & \gci\\
Spain & 2 & \gci\\ 
Switzerland & 2 & \thewu, \qswu, \edu \\
United Arab Emirates & 2 & \gci \\ 
Belgium & 1 & \thewu, \edu \\
Italy & 1 & \qswu \\
South Africa & 1 & \theee \\
Taiwan & 1 & \theee \\
\hline
\textbf{Total} & 101 &
\end{tabular}
\label{tab:countries}
\end{table}

The 101 universities offer only 45 programs with \enquote{cyber} in its name. The most prevalent are graduate (35), mostly master's. The rest are undergraduate (4) and other (6), such as post-baccalaureate or graduate certificates. The graduate programs typically last from 1 to 2 years, the undergraduate from 3 to 5.5 years, and others a few months. Most cyber programs overall are offered in the USA (17). Elsewhere, a maximum of three cybersecurity programs is offered.

The remaining 88 non-cyber programs include general computing programs, such as computer science, computer engineering, or software engineering. Some programs allow studying security or cybersecurity track or concentration. A few universities offer an information security program. 

From this point on, we will only focus on discussing the 45 cyber programs, since they best reflect the topic of this paper.

\subsection{Slow Adoption of Guidelines and Trends}

The data indicates that the top-ranked universities have not widely adopted the existing curricular guidelines. Even though cybersecurity is a distinct computing discipline, and there are specialized cybersecurity curricular frameworks (such as CSEC2017), many top universities do not provide any cybersecurity program, or offer only a certificate after passing a few semester-long courses.

Next, the majority of cyber programs do not relate themselves to the existing systematization initiatives, such as NSA CAE in Cybersecurity for curriculum and program requirements, or NIST NICE framework for describing cybersecurity tasks, knowledge, and skills. In particular, only 6 out of 17 cyber programs offered in the USA have been validated by NSA. However, the total number of NSA-designated institutions was 444 as of July 1, 2024~\cite{cae-map}. A good example of using external frameworks is the website of master program \textit{Cybersecurity} at New York University
~\cite{nyu-cybersecurity}, which presents the requirements for completing Cyber Defense and Cyber Operations tracks defined by NSA CAE in Cybersecurity.

There is also room for developing undergraduate programs. Our survey identified only four bachelor programs at top-ranked universities. 
However, three- or four-year-long programs enable more focus and specialization in cybersecurity than general computing programs or shorter master programs and certificates. The longer programs could also include more experiential learning, such as industry internships. 
A good example is the 4-year bachelor program in \textit{Cybersecurity Analytics and Operations} at Pennsylvania State University~\cite{penn-cyberanalytics}. This program contains dedicated cybersecurity courses, such as Cyber Incident Handling and Response, Cyber-Defense Studio, Malware Analytics, mandatory internship, and a cybersecurity capstone project.
There are other educational institutions that may provide this value through bachelor's programs -- in the USA alone, there are 178 programs available~\cite{cybersecurityguide} -- but often offered by other than top-ranked universities. Thus, they were not reflected in this survey due to the choice of the university rankings. 

\subsection{Inclusion of Non-technical Areas of Security}
\label{subsec:non-technical}

CSEC2017 curriculum guidelines recognize eight Knowledge Areas (KAs), which serve as the basic organizing structure for cybersecurity topics~\cite{csec2017}. Further, the guidelines list the essential concepts in all KAs that comprise the minimum required content for any cybersecurity program. In reference to these guidelines, \citet{blazic2022changing} argued that one of the ways to fulfill the demand for qualified professionals (cybersecurity skills gap) is the enrichment of the curricula with new content from the KAs that are covered the least, such as the organizational or human aspects of cybersecurity.

Therefore, we analyzed program websites and publicly available online materials to find out how the non-technical areas are represented. Namely, we searched for essential concepts defined in CSEC2017 for \emph{human, organizational, and societal security} in the descriptions of core or core elective courses.
If the concepts were only one or two topics out of ten or more in a single, typically introductory course, we did not count it.
Most often (25$\times$), the courses cover topics of organizational security, such as risk management, governance and policy, incident management, or security operations. 
Less often (12$\times$), they include societal security topics, for instance, cyber crime, cyber law, and cyber ethics.
The least represented (6$\times$) were topics of human security, for example, identity management, social engineering, or personal data privacy and security. 

Only five programs cover topics in the three KAs. One of these programs is a joint 2-year master \textit{Cybersecurity} at Estonian universities, Tallinn University of Technology and University of Tartu~\cite{taltech-mcs}. This program requires passing several semester-long courses dealing with the non-technical areas, such as Human Aspects of Cyber Security, Legal Aspects of Cyber Security, and Cyber Security Management. It also offers core electives, such as Privacy-preserving Technologies, Strategic Communications and Cybersecurity, Cyber Incident Handling, or Intelligence Methods for Cyber Professionals.

We believe that one of the reasons for the overall unsatisfactory situation is the lack of time in the master programs and the subsequent prioritization of technical KAs, such as data security, connection security, or system security. 
To address this gap, program directors at universities with faculties of law or social studies can consider enhancing their cybersecurity programs with relevant courses from these faculties. However, our experience from teaching at a university that offers a program with such courses shows that the course content needs to be adapted to be relevant to computing students and their future careers.

\subsection{Hands-on and Experiential Learning} \label{sec:hands-on}

CC2020 argues that \textit{competency} (as opposed to just knowledge) should be the standard for describing computing curricula. This can involve \enquote{stronger focus on various forms of experiential learning, from interactive simulations, to intensive projects, to field experiences, and to internships and cooperative programs with industry}~\cite{cc2020}. However, to fulfill this requirement, \enquote{domain-specific skills and dispositions require a learning environment that is different from a traditional classroom environment}~\cite{cc2020}. Accordingly, we examined whether the programs require completing internships in workplace-relevant settings and creating a final or capstone project or thesis that involves independent student work. 

\begin{figure}[!ht]

\centering
\includegraphics[width=0.9\linewidth]{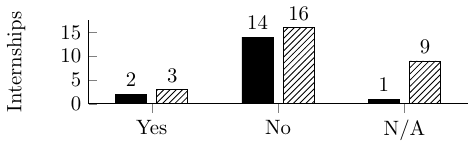}
\includegraphics[width=0.9\linewidth]{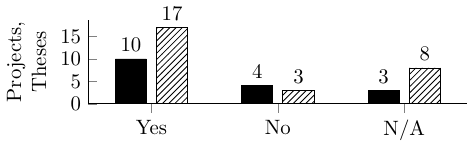}

\caption{Internships and final projects/theses for 17 US programs (black) and 28 outside the USA (hatch fill). \enquote{N/A} means the program descriptions do not contain this information.}

\Description{Internship distribution: (a) for 17 US programs: 2 yes, 14 no, 1 N/A; (b) for 28 non-US programs: 3 yes, 16 no, 9 N/A. Final project/thesis distribution: (a) for 17 US programs: 10 yes, 4 no, 3 N/A; (b) for 28 non-US programs: 17 yes, 3 no, 8 N/A.}
\label{fig:handson_analysis}
\end{figure}


\subsubsection{Internships}
\Cref{fig:handson_analysis} shows how many programs from the USA and outside require an internship.
The mandatory internships are present only in five programs (11\%). 
A notable example is bachelor \textit{Cybersecurity Engineering} at Tallinn University of Technology, which allocates a considerable number of credits (24 out of 180 ECTS credits) to the application of the gained skills and competencies in an authentic field-related working environment~\cite{taltech-bcs}.

However, including internships in a program assumes that all students will have enough opportunities to find a work placement that will be beneficial to their learning. This assumption might be difficult to fulfill for programs that have many students in regions with less-developed IT and cybersecurity ecosystems. Another challenge is the internship administration and assessment. 

\subsubsection{Projects and Theses}
The distribution of programs from the USA and outside that require a final project/thesis is shown in \Cref{fig:handson_analysis}.
The final project/thesis is mandated by 27 programs (60\%). The projects typically require independent student's work supervised by a faculty advisor on a cybersecurity topic. There are also projects where students interact with various organizations from a public, private, or nonprofit sector, solve their cybersecurity issues, and propose possible solutions.
Three programs include a mandatory semester-long or pre-capstone project. Completing both internship and final project/thesis is required by only four programs (9\%).

Regarding the written final thesis, we second to \citet{Perkins2024} who advocates for reducing the reliance on assessments where AI tools may be used to mimic human writing. Following the CC2020 recommendations, we suggest instead requiring the completion of an intensive practical project. However, projects bring other challenges, such as creating meaningful and workplace-relevant assignments and assessment of artifacts created by students.

\section{Cybersecurity Program Director's Checklist} \label{sec:checklist}

Based on the study findings and our experience, we created a 10-item checklist for directors of existing or prospective cybersecurity programs to initiate and support enhancements in their programs. 

\begin{enumerate}[itemsep=\smallskipamount]
    \item Is your \emph{program different} enough from other computing programs offered at your university? 
    \checklistitem{A dedicated cybersecurity program should include unique core courses not offered in other computing programs, ensuring specialized training for students.}

    \item Does your program cover \emph{all knowledge areas} defined by CSEC2017~\cite{csec2017}? Do you cover non-technical areas, such as \emph{legal and privacy aspects}? 
    \checklistitem{Since cybersecurity is an interdisciplinary field~\cite{csec2017}, consider involving specialists from non-computing areas, such as from other faculties or external experts. Subsequently, ensure their courses match the expected proficiency of the students.}

    \item Does your program teach \emph{specifics of national or regional regulations and laws}?
    \checklistitem{Even though cyberspace is global, nations regulate it differently~\cite{AlDaajeh2022}. Your graduates will benefit from courses covering the regulations relevant to their country or region. For instance, universities in the EU countries may teach directives valid for member countries and their local law~\cite{ecsf2022}.}
  
    \item Have you switched from the knowledge-based to the \emph{com-petency-\-based description of the curriculum} as suggested by CC2020~\cite{cc2020}?
    \checklistitem{Using \emph{competencies} instead of knowledge may encourage adopting new forms of experiential learning, as discussed in \Cref{sec:hands-on}. This can be a vital step to beat the skills gap~\cite{Crabb2024}.}
    
    \item How can you \emph{include internships}?
    \checklistitem{Since internships must be centered around cybersecurity, consider the appropriate credit allocation. Also, consider whether the surrounding cybersecurity ecosystem in your city or area can accommodate your students and still provide them with relevant mentoring and topics.}
    
    \item Do you offer several \emph{cybersecurity elective} courses?     
    \checklistitem{This enables students to specialize in a cybersecurity area of their choice. Such specialization would be valued by their future employers seeking that specific expertise~\cite{Graham2023, Jones2018, Ozyurt2024}.}
    
    \item Does your school have \emph{sufficient lab environment} to teach hands-on courses?
    \checklistitem{
    The practical activities often do not scale well, particularly in physical computer labs. 
    Online training platforms~\cite{Beuran2023} can address this issue with the benefit of being available to your students when needed. However, be aware that hands-on classes come with the cost of more laborious preparation and delivery on the instructors' side compared to onsite or online lectures and demonstrations.}
    
    \item Does your school have access to up-to-date \emph{training materials and environments} for hands-on classes? Can your instructors create and maintain these materials independently?     
    \checklistitem{Relevant teaching content that involves practice is key for programs that aim to fill the cybersecurity skills gap~\cite{Crabb2024}. Reusing and sharing training materials is challenging because classes are highly dependent on the technical environment. Still, instructors can leverage existing resources such as the CLARK Cybersecurity Library~\cite{clark}.}

    \item Have you considered \emph{changing the assessment} of courses, final projects, and thesis, given the spread of AI tools~\cite{Perkins2024}?
    \checklistitem{We believe that AI tools should be allowed in class and during assessments (exams, mid-term tests, and final projects) because practitioners use them as well. However, the assessments should move from simple question-and-answer formats to more complex assignments testing whether students can use concepts, tools, infrastructures, and data to produce useful output, thereby mimicking authentic workplace tasks~\cite{Ohm2024}.}
    
    \item Do you \emph{use existing national or international frameworks} when designing or describing the program or its courses?
    \checklistitem{These frameworks can serve as a common language across different worlds of universities, students, employers, and governments. For examples, see the list in \Cref{subsec:intro-cybersecurity}.}  
\end{enumerate}

\section{Conclusion} \label{sec:conclusion}

To answer the question posed in this paper's title, we examined how top-ranked world universities responded to the existing curricular guidelines (CSEC2017 and CC2020), which consider cybersecurity as a distinctive and interdisciplinary field.
Among 101 higher education institutions analyzed, we identified only 45 cybersecurity programs. 
It is concerning that out of these 45, only five cover all eight knowledge areas defined by CSEC2017, especially organization, societal, and human security. Other programs cover only areas traditionally taught in computer science or engineering programs or their security concentrations, such as data or system security. Next, internships that enable practicing competencies and dispositions as promoted by CC2020 are mandated only by five programs.
While more than half of the programs require a final project or thesis, six programs require neither an internship nor a final project/thesis. This lack of hands-on experience may leave graduates underprepared for their future careers. To support our claims, the dataset we collected is available as a supplementary material~\cite{dataset}.

\subsection{Recommendations for Future Improvements}

Our analysis revealed a fragmentation of information about study programs and courses. First, there is no global ranking specifically for universities that offer cybersecurity programs. The well-known QS World University Rankings or Times Higher Education World University Rankings do not recognize cybersecurity as a distinctive subject (as opposed to computer science, and most recently data science and AI). EduRank considers only research metrics, excluding teaching. As a result, prospective students and program directors cannot easily find and compare available program offerings. Moreover, the study program directors cannot use such rankings as a benchmark to initiate and foster improvements in their programs.

Further, descriptions of the program structure, content, and requirements are presented mostly in general language, with no reference to existing frameworks, such as CSEC2017 knowledge areas; NICE work roles, tasks, knowledge, and skills; or NSA CAE knowledge units. This is another barrier to comparing individual programs for both students and directors. Instead, universities could better leverage these frameworks to enable program comparison. Lastly, governments and higher education analytics companies could compile rankings focusing solely on cybersecurity programs.

\begin{acks}
This research has received funding from the European Union’s Horizon Europe program under Grant Agreement No. 101087529.
\end{acks}

\newpage    

\bibliographystyle{ACM-Reference-Format}
\balance
\bibliography{references}

\end{document}